\documentclass[11pt,journal,compsoc]{IEEEtran}

\ifCLASSOPTIONcompsoc
  \usepackage[nocompress]{cite}
\else
  \usepackage{cite}
\fi

\ifCLASSINFOpdf

\else

\fi

\usepackage{float,balance}
\usepackage{xcolor}
\usepackage{graphicx}
\usepackage{amsmath,epsfig,cite}
\usepackage{multirow}
\usepackage{epstopdf}
\usepackage{subfigure}
\usepackage{hyperref}
\usepackage{verbatim}
\usepackage{enumitem}
\usepackage{tikz}
\usepackage{cite}
\def \etal {et al.}

\begin{document}
%
\title{Unsupervised Personalization of an Emotion Recognition System: The Unique Properties of the Externalization of Valence in Speech \vspace{-0.3cm}}

\author{Kusha~Sridhar,~\IEEEmembership{Student Member,~IEEE,}
        and~Carlos~Busso,~\IEEEmembership{Senior member,~IEEE} \vspace{-0.3cm}
\IEEEcompsocitemizethanks{\IEEEcompsocthanksitem K. Sridhar and C. Busso are with the Erik Jonsson School of Engineering and Computer Science, The University of Texas at Dallas, ( E-mail: Kusha.Sridhar@utdallas.edu, busso@utdallas.edu)}}

\markboth{}%
{Sridhar \MakeLowercase{\textit{et al.}}: Personalizing an emotion recognition system through Valence}

\IEEEtitleabstractindextext{%
\begin{abstract}
The prediction of valence from speech is an important, but challenging problem. The externalization of valence in speech has speaker-dependent cues, which contribute to performances that are often significantly lower than the prediction of other emotional attributes such as arousal and dominance. A practical approach to improve valence prediction from speech is to adapt the models to the target speakers in the test set. Adapting a \emph{speech emotion recognition} (SER) system to a particular speaker is a hard problem, especially with \emph{deep neural networks} (DNNs), since it requires optimizing millions of parameters. This study proposes an unsupervised approach to address this problem by searching for speakers in the train set with similar acoustic patterns as the speaker in the test set. Speech samples from the selected speakers are used to create the adaptation set. This approach leverages transfer learning using pre-trained models, which are adapted with these speech samples. We propose three alternative adaptation strategies: unique speaker, oversampling and weighting approaches. These methods differ on the use of the adaptation set in the personalization of the valence models. The results demonstrate that a valence prediction model can be efficiently personalized with these unsupervised approaches, leading to relative improvements as high as 13.52\%. \vspace{-0.2cm}
\end{abstract}

}

\maketitle 

\IEEEdisplaynontitleabstractindextext

\IEEEpeerreviewmaketitle

\IEEEraisesectionheading{\section{Introduction}\label{sec:introduction}}

\IEEEPARstart{T}{he} 
area of \emph{speech emotion recognition} (SER) is an important research problem due to its key potential in fields such as \emph{human-computer interactions} (HCIs), healthcare \cite{Aldeneh_2019,Gideon_2019} and behavioral studies \cite{Narayanan_2013, Georgiou_2011}. Despite remarkable advances in emotion recognition, detecting emotions from speech is still a challenging task. The usual formulation to describe emotions is with categorical descriptors such as happiness, sadness, anger and neutral. However, this approach may not capture the intra and inter class variability across distinct emotional classes. 
{\color{black}
(i.e., variability across sentences with the same emotional class labels and variability across sentences with different emotional class labels).}
An alternative representation is the use of emotional attributes, as suggested by the core affect theory \cite{Russell_2003}. The most common attributes are arousal (calm versus active), valence (unpleasant versus pleasant) and dominance (weak versus strong). Because of their direct application in many areas, it is very important to build accurate models, which can reliably predict these emotional attributes. The estimation of emotional attributes is often posed as a regression problem, where the goal is to predict the scores associated with these attributes. In particular, the emotional attribute valence is key to understand many behavioral disorders {\color{black}\cite{groenewold2013emotional, hagenhoff2013reduced}} such as \emph{post-traumatic stress disorder} (PTSD), depression, schizophrenia and anxiety. Although different approaches have been proposed to improve SER systems, the prediction of valence using acoustic features is often less accurate than other emotional attributes such as arousal or dominance. The gap in performance is significant even with different methods specially implemented to correct this problem, such as using features from other modalities \cite{Nicolaou_2011,Aldeneh_2017_2,Zhang_2019_2,Tournier_2019}, modeling contextual information \cite{Mariooryad_2013_2} or regularizing \emph{deep neural network} (DNNs) under a \emph{multitask learning} (MTL) framework \cite{Parthasarathy_2017_3, Parthasarathy_2018_3}. It is important to explore why predicting valence from speech is so difficult, and use these findings and insights to improve SER systems.


In our previous work, we studied the prediction of valence from speech \cite{Sridhar_2018}, focusing the analysis on the role of regularization in DNNs. In particular, we explored the role of dropout as a form of regularization and analyzed its effect on the prediction of valence. Our analysis showed that a higher dropout rate (i.e., higher regularization) led to improvements in valence predictions. The optimal dropout rate for valence was higher than the optimal dropout rates for arousal and dominance across different configurations of the DNNs. A hypothesis from this study was that a heavily regularized network learns features that are more consistent across speakers, placing less emphasis on speaker-dependent emotional cues. We also conducted controlled speaker-dependent experiments to evaluate this hypothesis, where data from the same speakers were included in the train and test partitions.  
For valence, we observed relative gains in \emph{concordance correlation coefficient} (CCC) up to 30\% between speaker-dependent and speaker-independent experiments. The corresponding relative improvements observed for arousal and dominance were less than 4\%. These results showed that valence emotional cues include more speaker-dependent traits, explaining why heavily regularizing a DNN helps to learn more general emotional cues across speakers \cite{Sridhar_2018}. Building on these results, we propose an unsupervised personalization approach that is extremely useful in the prediction of valence. 

This paper explores the speaker-dependent nature of emotional cues in the externalization of valence. We hypothesize that a regression model trained to detect valence from speech can be adapted to a target speaker. The goal is to leverage the information from the emotional cues of speakers in the train set to fine-tune a regression model already trained to perform well on the prediction of valence. Our approach identifies speakers in the train set that are closer in the acoustic space to the speakers in the test set. Data from these selected speakers are used to create an adaptation set to personalize the SER models toward the test speakers. We achieve the adaptation by using three alternative methods: unique speaker, oversampling and weighting approaches. The unique speaker approach randomly selects samples from the data obtained from the selected speakers in the train set without replacement, regardless of how many times these speakers are selected (i.e., a speaker in the train set may be found to be closer to more than one speaker in the test set). 
{\color{black}The oversampling approach draws data from the selected speakers as a function of the number of times that a given speaker is selected. For instance, if a speaker in the train set is found to be closer to two speakers in the test set, the selected sentences from that training speaker is counted twice. This approach repeats the data from this speaker during the adaptation phase, so the model sees the same speech samples in multiple batches in a single epoch.} The weighting approach uses weights, where samples from the selected speakers in the train set are weighted more. This approach adds weights on the cost function during the training process, building the models from scratch. We demonstrate the idea of personalization under two scenarios: 1) separate SER models, where each of them is personalized to a single test speaker (i.e., individual adaptation models), and 2) a single SER model personalized to a pool of 50 target speakers (i.e., global adaptation model). We evaluate the approaches by monitoring the loss function on either a separate development set or the adaptation set.

Using the proposed model adaptation strategies leads to relative improvements in CCC as high as 13.52\% in the prediction of valence. 
While the adaptation experiments prove to be very effective for valence, the improvements achieved for arousal and dominance are less than 1.9\% (on the MSP-Podcast corpus). This result indicates the need for a personalization method to improve the prediction of valence, highlighting the benefits of our proposed approach. The contributions of our study are:

\begin{itemize}[leftmargin=0em]
\vspace{-0.2em}
\setlength{\itemindent}{1em}
\setlength{\itemsep}{0cm}%
\setlength{\parskip}{0cm}%
\item {\color{black}We leverage the finding that the externalization of valence in acoustic features is more speaker-dependent than arousal and dominance, raising awareness on the need for special considerations in its detection.} 
\item We successfully personalize a SER system using unsupervised adaptation strategies by exploiting the speaker-dependent traits.
\item We propose three alternative adaptation strategies to personalize a SER system, obtaining important relative performance improvements in the prediction of valence.
\end{itemize}

{\color{black}
One of the key strengths of this study is that we find similar speaker in the emotional feature space alone. By exploiting similarities in the emotional feature space, we have suppressed the speaker trait or text dependencies. Our approach provides a much more powerful way of comparing emotional similarities than traditional methods used for speaker identification. Likewise, our personalization approach avoids or minimizes ``concept drift.'' SER is a challenging problem, where the prediction models can become more volatile with the addition of more data over time. If the distribution of the newly acquired data starts to diverge or tend to fill up the sparse regions of the old data's distribution, the prediction results may see a drop in performance. Therefore, models built for analyzing such data quickly become obsolete. This phenomenon is referred to as concept drift. With our personalization study, we can minimize the impact of concept drift by developing personalized SER models that are tailored to target speakers. We can periodically re-fit or update the models to target speakers or even weight the data based on their historical significance to develop better personalized models.}

The paper is organized as follows. Section \ref{sec:related} discusses relevant studies on the prediction of valence from speech. It also describes the adaptation and personalization approach proposed for improving SER systems. Section \ref{sec:resources} presents the database used in this study. Section \ref{sec:analysis_1} describes the analysis on the role of regularization in the prediction of valence from speech, summarizing the study presented in our preliminary work \cite{Sridhar_2018}. Section \ref{sec:personalization} presents the proposed formulation to personalize a SER system, building on the insights learned  from the analysis in Section \ref{sec:analysis_1}. Section \ref{sec:results} presents the results obtained by using our proposed approaches to personalize a SER system. {\color{black}We primarily present the results on the MSP-Podcast corpus, but we evaluate the generalization of our proposed approach with two other databases.} The paper concludes with Section \ref{sec:conclusion}, which summarizes our key findings, providing future directions for this study.

\vspace{-0.3cm}
\section{Related Work}
\label{sec:related}

\subsection{Improving the Prediction of Valence}
\label{ssec:valence_importance}

While valence is a key dimension to understand complex human behaviors, its predictions using speech features are often lower than the predictions of other emotional attributes such as arousal or dominance \cite{Trigeorgis_2016}. Therefore, several speech studies have focused on understanding and improving valence prediction. 
Busso and Rahman \cite{Busso_2012} studied acoustic properties of emotional cues that describe valence. They built separate \emph{support vector regression} (SVR) models trained with different groups of acoustic features: energy, fundamental frequency, voice quality, spectral, \emph{Mel-frequency cepstral coefficients} (MFCCs) and RASTA features. They also built binary classifiers to distinguish between two groups of sentences characterized by similar arousal but different valence. The study showed that spectral and fundamental frequency features are the most discriminative for valence. Koolagudi and Rao \cite{Koolagudi_2009} claimed that MFCCs were effective to classify emotion along the valence dimension (i.e., spectral features). Cook \etal \cite{Cook_2005, Cook_2006} explored the structure of the fundamental frequency (F0), extracting dominant pitches in the detection of valence from speech. Despande \etal \cite{Deshpande_2019} proposed a reduced feature set consisting of the autocorrelation of pitch contour, \emph{root mean square} (RMS) energy and a 10- dimensional \emph{time domain difference} (TDD) vector. The TDD vector corresponds to successive differences in the speech signal.  The feature set collectively led to better results than MFCCs or OpenSmile features \cite{Deshpande_2019_2}. Tursunov \etal \cite{Tursunov_2019} used acoustic descriptors associated with timbre perception to classify discrete emotions, and emotions along the valence dimension. Tahon \etal \cite{Tahon_2012} showed that voice quality features were also useful in the detection of valence. 

Other studies have explored modeling strategies to improve the prediction of valence. Lee \etal \cite{Lee_2009_2} used dynamic Bayesian networks to capture time dependencies and mutual influence of interlocutors during dyadic interactions. Contextual information was found to be particularly useful in the prediction of valence, leading to relative improvements higher than the one observed for arousal. Another alternative approach to improve valence was by regularizing a DNN. 
For example, Parthasarathy and Busso \cite{Parthasarathy_2017_3} showed that jointly predicting valence, arousal and dominance under a \emph{multitask learning} (MTL) framework helps to improve its prediction. The MTL framework acts as regularization in DNNs. Other approaches using MTL have shown similar findings. Since the performance of lexical models often outperforms acoustic models in predicting valence \cite{Aldeneh_2017_2}, Lakomkin \etal \cite{Lakomkin_2019} suggested the use of the output of an \emph{automatic speech recognition} (ASR) as the input of a character-based DNN.

\vspace{-0.3cm}
\subsection{Model Adaptation in SER Tasks}
\label{ssec:adapting_SER}

Unlike other speech tasks such as \emph{automatic speech recognition} (ASR) that rely on abundant data, databases used in SER are often small. Therefore, many researchers have explored the use of model adaptation techniques to generalize the models beyond the training conditions. Most of the adaptation techniques aim to attenuate sources of variability including channel, language and speaker mismatches. Early studies demonstrated the effectiveness of these techniques with algorithms based on \emph{support vector machine} (SVM) \cite{Abdelwahab_2015}. 
Abdelwahab and Busso \cite{Abdelwahab_2017_2} demonstrated the importance of data selection strategy for domain adaptation. They illustrated that, incrementally adapting emotion classification models using active learning to select samples from the target domain can improve their performance. They used a conservative approach where only the correctly classified samples were used to adapt the model, leaving out the incorrect ones in order to avoid large changes in the hyperplane between the classes.

Recent efforts in model adaptation have mainly focused on DNNs, where important advances have been made in the area of transferring knowledge between domains \cite{Bengio_2011}. DNNs with their deep architectures can learn useful representations by compactly representing functions. Deng \etal \cite{Deng_2013} used sparse autoencoders to learn feature representations in the source domain that are more consistent with the target domain. This goal was achieved by simultaneously minimizing the reconstruction error in both domains. Deng \etal \cite{Deng_2014} proposed the use of unlabeled data under a deep autoencoder framework to reduce the mismatch between train and test conditions. They also simultaneously learned common traits from both labeled and unlabeled data. 

Instead of the traditional method of pre-training and fine-tuning for model adaptation, Gideon \etal \cite{Gideon_2017} used progressive networks to enhance a SER system. They trained the model on new tasks by freezing the layers related to previously learned tasks and used their intermediate representations as inputs to new parallel layers. This study also used paralinguistic information from gender and speaker identity  to achieve improvements. Similarly, other variants of adaptation techniques use \emph{kernel mean matching} (KMM) \cite{Hassan_2013}, \emph{Nonnegative matrix factorization} \cite{Song_2016}, \emph{domain adaptive least-squares regression} \cite{Zong_2016}, and PCANet \cite{Huang_2017}. These methods lead to improvements on emotion recognition tasks, by using hybrid frameworks involving unsupervised followed by supervised learning. Our proposed approach is different from these studies, since we aim to explicitly exploit similarities between speakers in the train and test sets, as measured in the feature space. This approach leads to powerful adaptation methods that are particularly useful to predict valence. 

\vspace{-0.3cm}
\subsection{Speech Emotion  Personalization}
\label{ssec:personalize_SER}

This study focuses on adapting or personalizing a SER system to a target set of speakers. Busso and Rahman \cite{Rahman_2012} demonstrated the idea of personalization using an unsupervised feature normalization scheme. They used the \emph{iterative feature normalization} (IFN) method \cite{Busso_2013_2} to reduce speaker variability, while preserving the discriminative information of the features across emotional classes. The IFN algorithm has two steps. First, it detects neutral sentences which are used to estimate the normalization parameters. Then, the data is normalized with these parameters. Since the detection of neutral speech is not perfect, this process is iteratively repeated leading to important improvements. Busso and Rahman \cite{Rahman_2012} implemented the IFN scheme as a front end of a SER system designed to recognize emotion from a target speaker, observing huge improvements in accuracy. Our study exploits the speaker-dependencies in the externalization of valence to personalize a SER system towards target speakers.

\vspace{-0.3cm}
\section{Resources}
\label{sec:resources}

\subsection{Emotional Corpora}
\label{ssec:corpora}

\subsubsection{The MSP-Podcast Corpus}
\label{sssec:corpus}

The study relies on the MSP-Podcast corpus \cite{Lotfian_2019_3}, which provides a diverse collection of spontaneous speech segments that are rich in emotional content. The speech segments are obtained from podcasts taken from various audio-sharing websites, using the retrieval-based approach proposed by Mariooryad \etal \cite{Mariooryad_2014_3}. The content of the podcasts is diverse, including discussions on sport, politics, entertainment, games, social problems and healthcare. The podcasts are segmented into speaking turns between 2.75s and 11s duration. These segments are automatically processed to discard segments with music, overlapped speech, and noisy recordings. Since most of the segments are expected to be neutral, we retrieve candidate segments to be included in the database by leveraging a diversified set of SER algorithms to detect emotions. The selected speech segments are annotated on \emph{Amazon Mechanical Turk} (AMT) using a crowdsourcing protocol similar to the one introduced by Burmania \etal  \cite{Burmania_2016_2}. This crowdsourcing protocol stops the annotators in real-time if their performance is evaluated as poor. The raters annotate each speaking turn for its arousal, valence and dominance content using \emph{self-assessment manikins} (SAMs) on a seven Likert-type scale. The ground truth labels for each speaking turn is the average across the scores provided by the annotators. Although we do not use categorical annotations in this study, the corpus also includes annotations of primary and secondary emotions. The primary emotion corresponds to the dominant emotional class. The secondary emotion corresponds to all the emotional classes that can be perceived in the speech segments. 

\begin{figure}[tb]
	\centering
	\includegraphics[width=0.98\columnwidth]{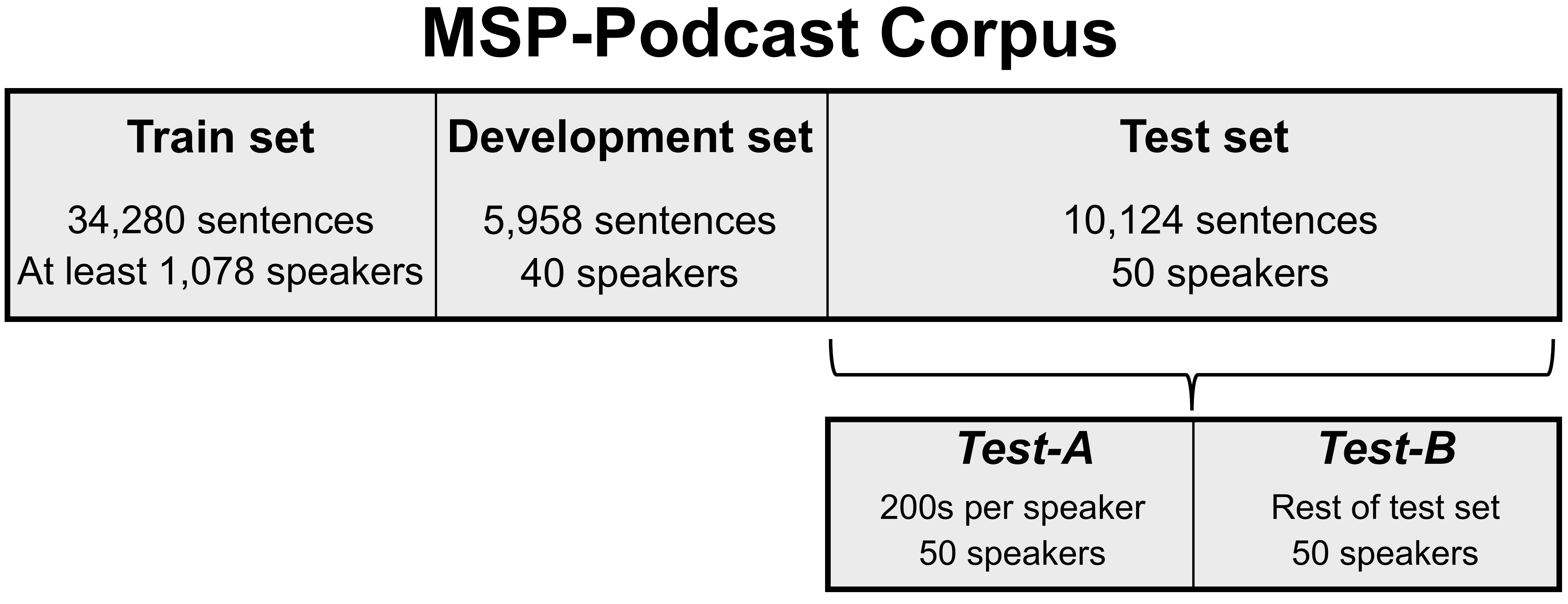}
	\caption{Partitions of the MSP-Podcast corpus used in this study for the train, development and test sets. The test set is further split into the \emph{test-A} and \emph{test-B} sets.}
	\vspace{-0.3cm}
	\label{fig:corpus}
\end{figure}

{\color{black}
The collection of the MSP-Podcast corpus is an ongoing effort. This study uses version 1.6 of the MSP-Podcast corpus, which consists of 50,362 speech segments (83h29m) annotated with emotional classes. From this set, 42,567 segments have been manually assigned to 1,078 speakers. The speaker identity for the rest of the corpus has not been assigned yet. Figure \ref{fig:corpus} illustrates the partition of the dataset used in this study. The test set has 10,124 speech segments from 50 speakers, and the development set has 5,958 speech segments from 40 speakers. Each speaker in the test and development sets has a minimum of five minutes of data. The rest of the corpus is included in the train set, which consists of a total of 34,280 speech segments. The data partition aims to create speaker-independent partitions between sets. Lotfian and Busso \cite{Lotfian_2019_3} provide more details on this corpus.
}

As shown in Figure \ref{fig:corpus}, we further split the test set into two partitions for this study: \emph{test-A} and \emph{test-B} sets. The  \emph{test-A} set includes 200s of recording for each of the 50 speakers in the test set. The \emph{test-B} set includes the rest of the recordings in the test set. {\color{black}Each test speaker has at least 300s (5 mins) of data. After removing 200s from each speaker to form the \emph{test-A} set, the \emph{test-B} set is left with at least 100s of data for each speaker. The average duration per speaker in the \emph{test-B} set is 1005.96s}.

{\color{black}
\subsubsection{The IEMOCAP and MSP-IMPROV Corpora}
\label{ssec:corpus_2}

Besides the MSP-Podcast corpus, we use two other databases for our experimental evaluations. The first database is the USC-IEMOCAP corpus \cite{Busso_2008_5}, which is an audiovisual corpus and contains dyadic interactions from 10 actors in improvised scenarios. This study only uses the audio. The database contains 10,039 speaking turns, which are annotated with emotional labels for arousal, valence and dominance by at least two raters using a five-Likert scale. We also use the MSP-IMPROV corpus \cite{Busso_2017}, which is a multimodal emotional database that contains interactions between pairs of actors engaged in improvised scenarios. In addition to the conversations during improvised scenarios, the dataset also contains the interactions between the actors during the breaks, resulting in more naturalistic data. The corpus uses a novel elicitation scheme, where two
actors in an improvised scenario leads one of them to utter target sentences. For each of the target sentences, four emotional scenarios were created to contextualize the sentence to elicit happy, angry, sad and neutral reactions, respectively. This corpus consists of 8,438 turns of emotional sentences recorded from 12 actors (over 9 hours). The sessions were manually segmented into speaking turns, which were annotated with emotional labels using perceptual evaluations. Each turn was annotated for arousal, valence and dominance by five or more raters using a five-Likert scale. In both databases, the consensus emotional attribute label assigned to each utterance is the average across the scores provided by the annotators, which is linearly mapped between -3 and 3.}

\vspace{-0.3cm}
\subsection{Acoustic Features}
\label{ssec:features}

This study uses the feature set proposed for the \emph{computational paralinguistics challenge} (ComParE) in Interspeech 2013 \cite{Schuller_2013}. The features are extracted by estimating several \emph{low-level descriptors} (LLDs) such as energy, fundamental frequency and MFCCs. For each speech segment, statistics such as mean, standard deviation, range and regression coefficients are estimated for each LLD, creating \emph{high-level descriptors} (HLDs). With this approach, the feature vector is fixed regardless of the duration of the sentence. The ComParE set creates a 6,373 dimensional feature vector for each sentence.

\vspace{-0.3cm}
\section{Role of Regularization}
\label{sec:analysis_1}

Our proposed personalization method for valence builds upon the findings reported in our preliminary study \cite{Sridhar_2018}. This section summarizes the main findings on the role of dropout rate as a form of regularization in DNNs and its impact on SER.  The study in Sridhar \etal \cite{Sridhar_2018} was conducted on an early version of this corpus. We update the analysis with the release of the corpus used for this study (release 1.6 of the MSP-Podcast corpus). 

\vspace{-0.3cm}
\subsection{Optimal Dropout Rate for Best Performance}
\label{ssec:dropout_rate}

\begin{figure}[t]
	\includegraphics[width=0.90\columnwidth]{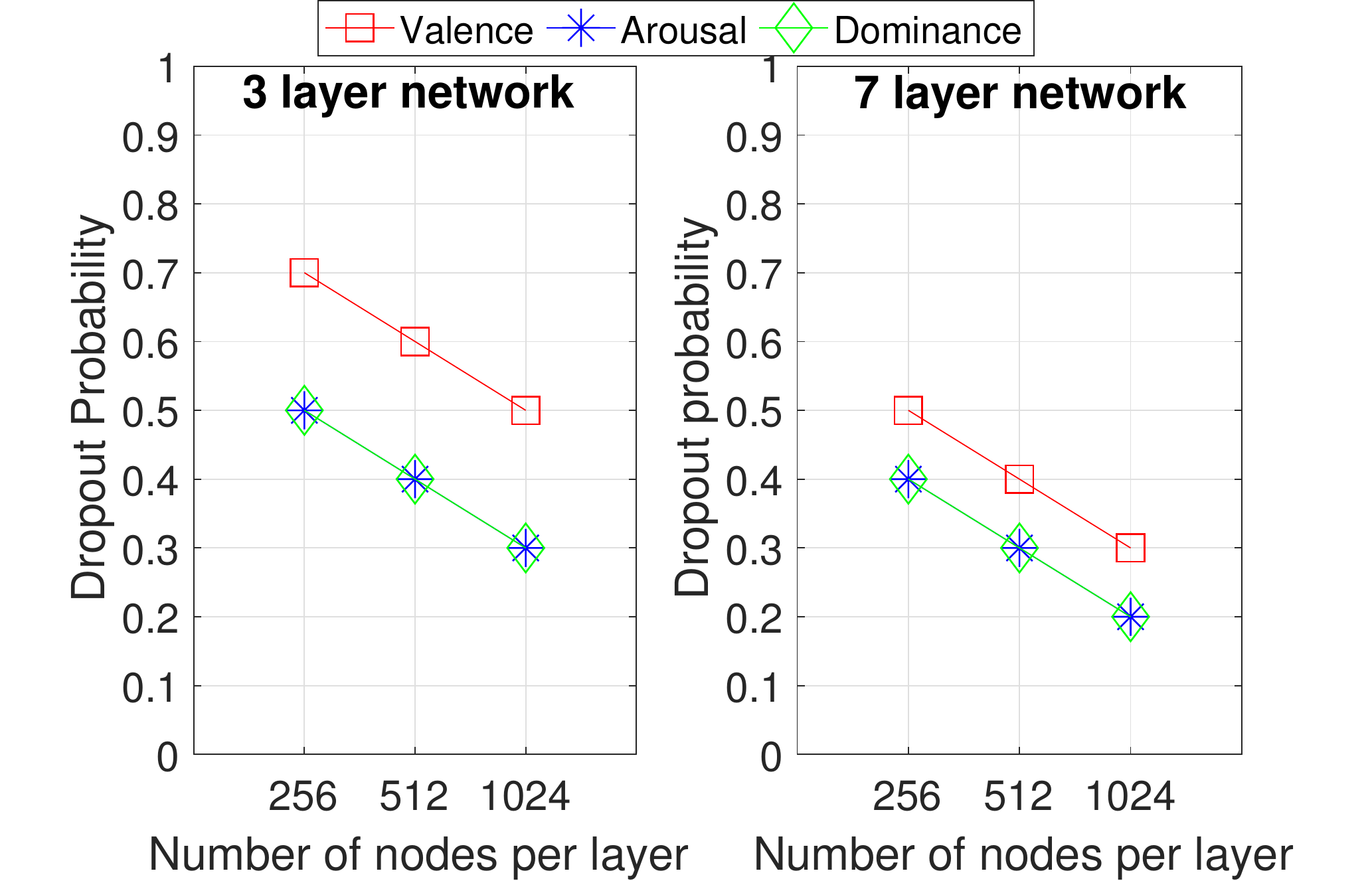}
	\vspace{-0.3cm}
	\caption{Optimal dropout rate observed in the development set as a function of the number of nodes per layer in a DNN. The DNN is implemented with either three or seven layers.}
	\label{fig:nodes_vs_dropout}
	\vspace{-0.3cm}
\end{figure}

Our previous study focused on the role of dropout as a form of regularization in improving the prediction of valence \cite{Sridhar_2018}. When dropout is used in DNNs, random portions of the network are shutdown at every iteration, training a smaller network on each epoch. This approach helps in learning feature weights in random conjunction of neurons, preventing developing co-dependencies with neighboring nodes. This regularization approach leads to better generalization. We explore the role of regularization in the prediction of valence by changing the dropout rate $p$. {\color{black}The goal of this analysis is to understand the optimal value $p$ that leads to the best performance for different network configurations (i.e., different number of layers, different number of nodes per layer). We train the models for 1,000 epochs, with an early stopping criterion based on the development loss. The loss function is based on CCC, which has led to better performance than \emph{mean squared error} (MSE) \cite{Trigeorgis_2016}. We train separate regression models by changing the dropout rate  $p\in\{ 0.0, 0.1, \cdots, 0.9\}$, recording the optimal dropout rate leading to the best performance on the development set. We evaluate two networks with three and seven layers, implemented with 256, 512 and 1,024 nodes per layer. Figure \ref{fig:nodes_vs_dropout} illustrates the results, showing the optimal dropout rate observed in the development set. The optimal dropout rates that give the best performance are higher for valence than arousal and dominance. While the optimal dropout rate decreases as we increase the number of layers or number of nodes per layer in a DNN, Figure \ref{fig:nodes_vs_dropout} shows that the gap between the optimal dropout rates for valence and arousal/dominance stays consistent. Interestingly, the optimal dropout rates for arousal and dominance are exactly the same across different DNN configurations, whereas it is different for valence. The results show that the need for higher regularization for valence is consistent across variations in the architectures of the DNNs.} 

\vspace{-0.3cm}
\subsection{Speaker-Dependent versus Speaker-Independent Models}
\label{ssec:sp_dep}

Section \ref{ssec:dropout_rate} demonstrated that a DNN needs to be heavily regularized to give good predictions for valence. We hypothesize that this finding can be explained by the speaker-dependent nature of speech cues in the externalization of valence (i.e., we use different acoustic cues to express valence). A DNN with higher regularization, learns more generic trends present across speakers, leading to better generalization. To validate our hypothesis, we conduct a controlled emotion detection evaluation, where we train DNNs with either speaker-dependent or speaker-independent partitions. SER should be performed with speaker-independent partitions, where the data in the train and test partitions are from disjoint set of speakers. A model trained with data from speakers in the test set has an unfair advantage over a system evaluated with data from new speakers, resulting in overestimated performance. Our goal is to quantify the benefits of using speaker-dependent partitions. 

\begin{table}[t]		
	\caption{Comparison of CCC values between speaker-independent and speaker-dependent conditions. The DNN is trained with four layers. The column `Gain' shows the relative improvement by training with partial data from the target speakers (\emph{test-A} set).}
	\centering
    \fontsize{8}{9}\selectfont
	\begin{tabular*}{1\columnwidth}{@{\extracolsep{\fill}}c|c|c|c|c}
		\hline
		 Attributes & Nodes & Speaker & Speaker & Gain\\
		 &&Independent&Dependent\\
		 \cline{3-5}
		 && \emph{Test-B} set& \emph{Test-B} & (\%)\\
		\hline
		\hline
		\multirow{3}{*}{Valence}
        & 256 & 0.3076 & 0.3373 & 9.65\\
		& 512 & 0.3083 & 0.3670 & 19.03\\        
		& 1,024 & 0.2997 & 0.3538 & 18.05\\        
		\hline
        \multirow{3}{*}{Arousal}
        & 256 & 0.7153 & 0.7216 & 0.88\\
        & 512 & 0.7164 & 0.7331 & 2.33\\
        & 1,024 & 0.7104  & 0.7258 & 2.16\\
		\hline
		\multirow{3}{*}{Dominance}
        & 256 & 0.6300 & 0.6379 & 1.25\\
        & 512 & 0.6374 & 0.6565 & 2.99\\
        & 1,024 & 0.6253 & 0.6352 & 1.58\\
		\hline
	\end{tabular*}
	\label{tab:results_within2}
	\vspace{-0.3cm}
\end{table}

We build DNNs with four layers, implemented with 256, 512 or 1,024 nodes. The speaker-dependent model is built by adding the \emph{test-A} set in the train set (Fig. \ref{fig:corpus}). This approach creates a train set with partial knowledge about the speakers in the test set. In contrast, the speaker-independent model uses only the train set. To have a fair comparison, both models are evaluated on the \emph{test-B} set with speech samples that are not used to either train or optimize the parameters of the systems. Table \ref{tab:results_within2} shows the CCC values of the models for speaker-independent and  speaker-dependent conditions. The last column calculates the relative improvements achieved under the speaker-dependent condition. We observe a performance gain up to 19.03\% for valence. The performances for the arousal and dominance models also increase, but the relative improvements are less than {\color{black}3\%. The fact that the performances increase using speaker dependent sets is expected. What is unexpected is that the relative gain is significantly higher for valence than for arousal and dominance.} These results clearly show that learning emotional traits from the target speakers in the test set has clear benefits for valence, validating our hypothesis that the externalization of valence in speech has speaker-dependent traits.


\vspace{-0.3cm}
\section{Proposed Personalization Method}
\label{sec:personalization}

\subsection{Motivation}
\label{ssec:method}

The findings in Section \ref{sec:analysis_1} suggest that leveraging data from speakers in the train set that are \emph{closer} to our target speakers in the test set should benefit our SER models. This is the premise of our proposed approach. We aim to improve the prediction of valence, bridging the gap in performance between speaker-dependent and speaker-independent conditions reported in Table \ref{tab:results_within2}. {\color{black} Data sampled from the selected speakers' recordings are used to create an adaptation set, as illustrated in Figure \ref{fig:approach_a}}. Once the closest speakers are identified, we can either adapt the models or assign more weights to samples from this adapatation set. This section describes our unsupervised personalization approach to improve the prediction of valence. Unlike the speaker-dependent settings used in Section \ref{ssec:dropout_rate} (and Sec. \ref{ssec:iemocap_improv}), the analysis and experiments in this study operate with speaker-independent partitions for train, development and test sets. The assumption in our formulation is that we have the speaker identity associated with each sentence in the test set.

\vspace{-0.3cm}
\subsection{Estimation of Similarity Between Speakers}
\label{ssec:estimation_close_speakers}

A key step in our approach is to identify speakers in the train set that are \emph{closer} to the speakers in the test set. Ideally, we would like to identify speakers who externalize valence cues in speech in a similar way. This aim is difficult with no clear solutions. We simplify our formulation by searching for similarities between speakers in the space of emotional speech features. {\color{black}By exploiting similarities on the emotional feature space, we expect to focus more on emotional patterns than on speaker traits, which would be the focus of speaker embeddings created with methods such as i-vector \cite{Dehak_2011} or the x-vector \cite{Snyder_2018}}. Our approach relies on \emph{principal component analysis} (PCA) to reduce the dimension of the space, followed by fitting a \emph{Gaussian mixture model} (GMM) to the resulting reduced feature space.

We aim to quantify the similarity in the feature space between the speaker $i$ in the train set, and the speaker $j$ in the test set, $d(i,j)$. The first step is to reduce the feature space, since we consider a high dimensional feature vector (6,373D -- Sec. \ref{ssec:features}). Reducing the feature space creates a more compact feature representation, where the similarity between speakers can be more efficiently computed. We implement this step with PCA, which is a popular unsupervised dimensionality reduction technique. {\color{black}First, we estimate the zero-mean vector $\mathbf{y}_s=\mathbf{f}_s-\mathbf{\bar{f}}$, where $\mathbf{f}_s$ is the feature vector of sentence $s$, and $\mathbf{\bar{f}}$ is the mean feature vector. Then, we concatenate these $M$ vectors, creating matrix $F$ (Eq. \ref{eq:conc}). From this matrix, we estimate the sample covariance matrix $Q$ using Equation \ref{eq:sample}. Then, we compute the eigenvectors of $Q$, selecting the ones with the highest eigenvalues, which are considered as the \emph{principal components} (PCs).  

\vspace{-0.3cm}
\begin{eqnarray}
F&=& [\mathbf{y}_1,\mathbf{y}_2,\ldots,\mathbf{y}_M] \label{eq:conc}\\
Q&=& \frac{1}{M-1} FF^T \label{eq:sample}
\end{eqnarray}
}
\vspace{-0.3cm}

The PCA-based feature reduction is implemented for each speaker in the test set, creating speaker-dependent transformations. {\color{black}We use the 10 most important dimensions, which explain in average 57.9\% of the variance in the feature space.} The speech sentences from speaker $i$ (train set) are projected into the PCA space associated with speaker $j$ (test set). The speech sentences from speaker $j$ are also projected in that space. After the PCA projections, we fit two separate GMMs on the reduced feature space, one for the sentences of speakers $i$ ($p_{i}^{\mathit{train}}$), and another for the sentences of speaker $j$ ($q_{j}^\mathit{test}$). The GMMs have 10 mixtures, matching the reduced dimension of the PCA projections. 
Finally, we estimate the similarity between the GMMs using the \emph{Kulback Liebler Divergence} (KLD). 

\vspace{-0.3cm}
\begin{eqnarray}
p_{i}^{\mathit{train}} (x_i)&=&\sum_{n=1}^{10} w_{n(i)} N(x_i,\mu_{n(i)},\Sigma_{n(i)})\\
q_{j}^\mathit{test} (x_j)&=&\sum_{n=1}^{10} w_{n(j)} N(x_j,\mu_{n(j)},\Sigma_{n(j)})\\
d(i,j) &=& KLD (p_{i}^{\mathit{train}}, q_{j}^{\mathit{test}})
\end{eqnarray}

\begin{figure}[tb]
	\centering
	\subfigure[Selection of closest speakers to create adaptation set]{
		\includegraphics[width=0.98\columnwidth]{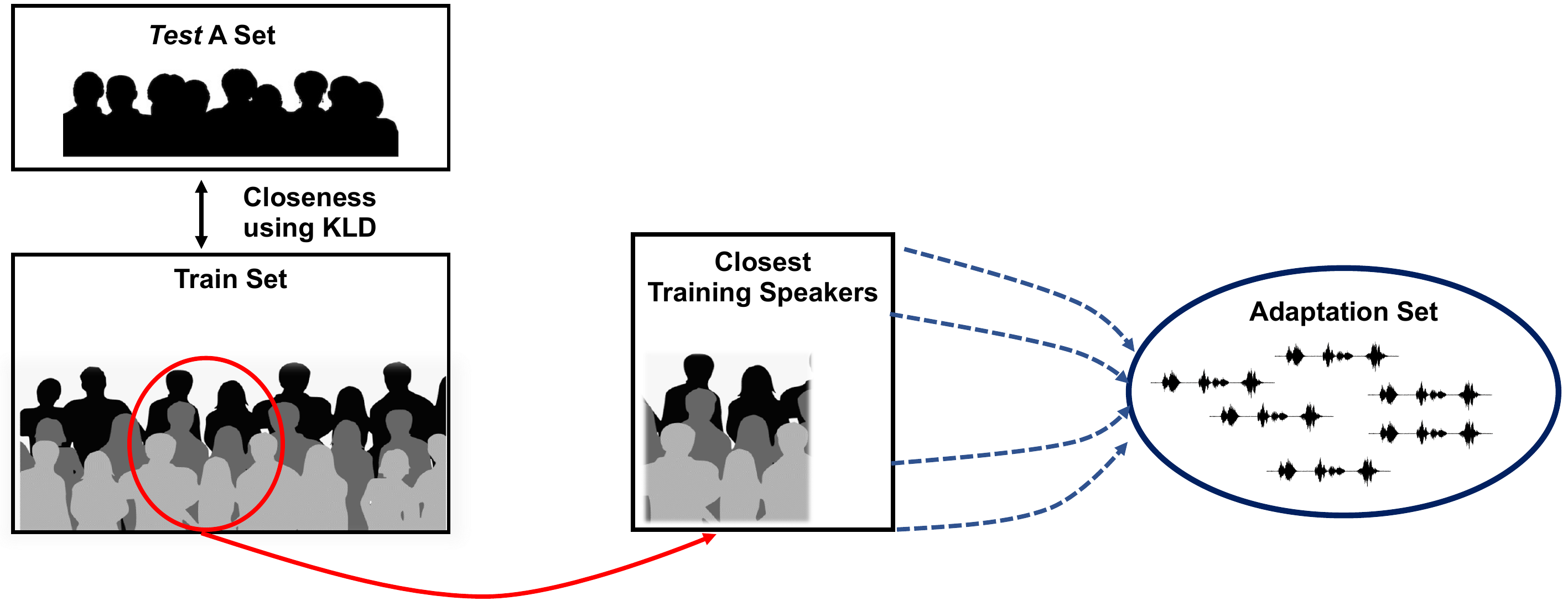}
		\label{fig:approach_a}
	}
	\subfigure[Criteria for selecting samples for the adaptation set]{
		\includegraphics[width=0.98\columnwidth]{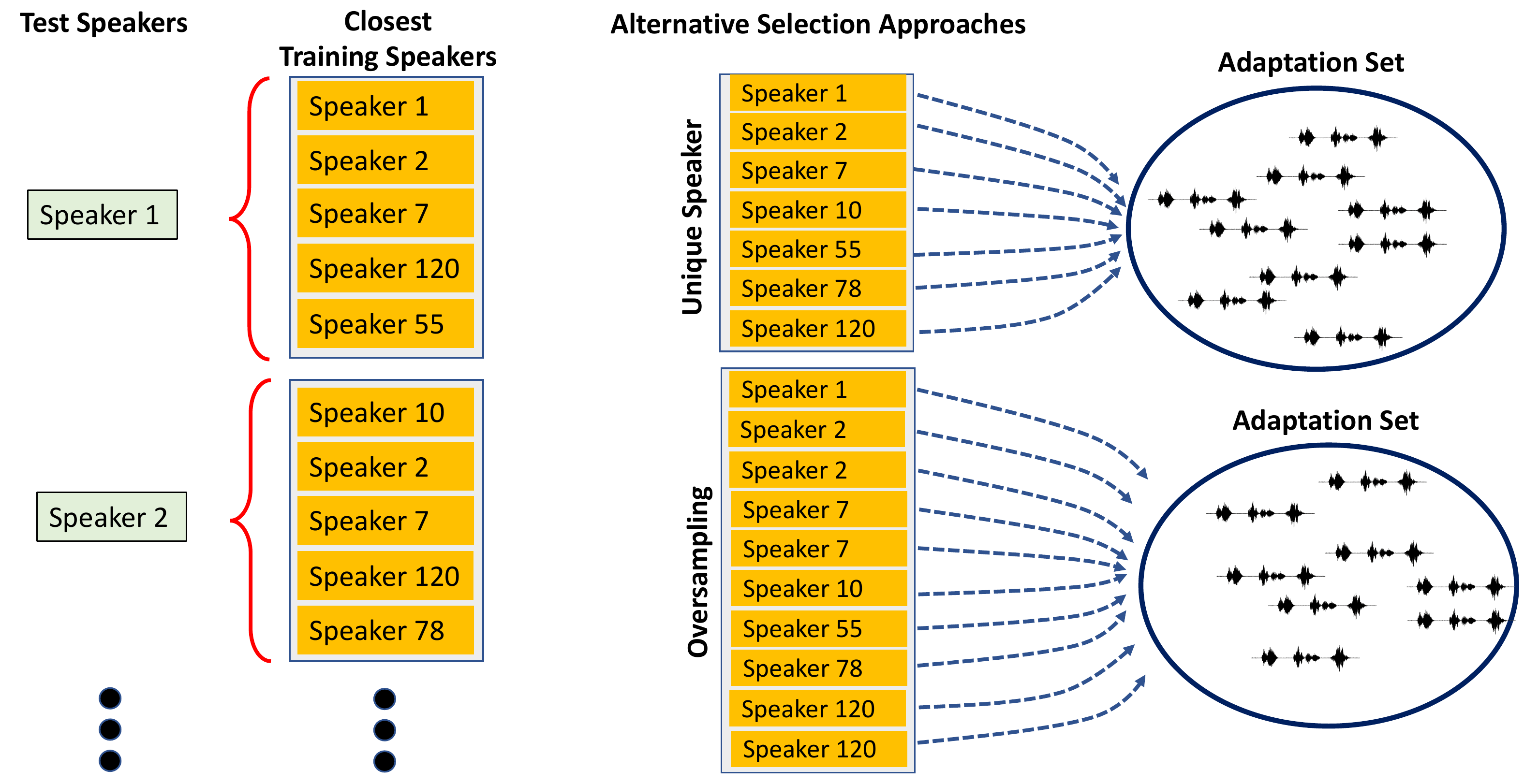}
		\label{fig:approach_b}
	}
	\caption{Illustration of proposed personalization approach. We identify the closest set of speakers in the train set to each of the target speaker in the test set. Sentences from these speakers are randomly sampled with different selection criteria.}
	\label{fig:approach}
	\vspace{-0.3cm}
\end{figure}

For a given speaker $j$ in the test set, we estimate $d(i,j)$ for all the speakers in the train set, sorting their scores in increasing order. The closest speakers in the train set are the top speakers in this ranked list. This approach is repeated for each of the 50 speakers in the test set {\color{black}(i.e., we have 50 different PCA projections)}. While this step can be implemented using all the data from the test set, we use the \emph{test-A} set to have the same amount of data for each speaker (i.e., 200s -- Fig. \ref{fig:corpus}).  {\color{black}Figure \ref{fig:approach_a} illustrates the process to form the adaptation set by finding the closest set of training speakers to a target speaker. Notice that the adaptation set is a subset of the train set, for which we  have labels.}

\vspace{-0.3cm}
\subsection{Personalization Approach}
\label{ssec:Personalization}

After selecting the speakers in the train set that are closest to each of the test speakers, our next step is to leverage data from these train set speakers to either personalize or adapt the emotion prediction models for valence. We propose and evaluate three alternative methods, referred to as \emph{unique speaker}, \emph{oversampling}, and \emph{weighting} approaches. The first two approaches rely on adapting a model. We built a regression model using the train set to predict valence. The weights of the pre-trained models are frozen with the exception of the last layer, which are fine-tuned using the adaptation data. The last approach requires training the network from scratch. 

\noindent
\underline{Unique speaker approach}: {\color{black}
Each speaker in the test set generates a list with its $N$ closest speakers in the train set. Since some speakers in the train set may be close to more than one speaker, the total number of selected speakers after combining the lists from the 50 speakers in the test set is less or equal to $N\times50$. The unique speaker approach considers all the data from these speakers in the train set. We create the adaptation set by sampling from the data from these speakers without replacement. Therefore, each speech segment can be considered only once in the adaptation set. Figure \ref{fig:approach_b} illustrates the process for the case when we have only 2 speakers in the test set. In the example, speakers 2, 7 and 120 in the train set are found to be close to both test speakers, hence, we consider them only once when forming the adaptation set. We implement a balance sampling criterion that aims to select approximately the same amount of data for each speakers. For example, for an adaptation set of 200s, if we have 7 unique speakers selected as the closest speakers from the test speakers, as in the example, we would randomly select approximately 28.6s for each of these speakers. We adopt this approach in an attempt to diversify and balance the speech samples selected from all the speakers in the unique speakers set. This approach uses the pre-trained models trained with all the train set, personalizing the models with the adaptation set.} 

\noindent
\underline {Oversampling approach}: {\color{black}A speaker in the train set may be in the list of the closest speakers for more than one speaker in the test set. The oversampling approach assumes that these samples are more relevant during the adaptation process. If a speaker is selected $C$ times (i.e., the speaker is one of the closest speakers for $C$ speakers in the test set), the oversampling method will create $C$ copies of his/her sentences before randomly drawing the samples. This process is illustrated in Figure \ref{fig:approach_b} were speakers 2, 7 and 120 in the train set are copied twice. Therefore, more samples from these speakers will appear in the adaptation set. We form the adaptation set with a balance approach, choosing approximately the same amount of data from the selected speakers. In the example from Figure \ref{fig:approach_b}, we would select 20 second for each of the 10 sets for an adaptation set of 200s. Sentences can even be repeated on the adaptation set. This approach also fine-tunes the pre-trained model using speech samples from the oversampled adaptation set.}

\noindent
\underline{Weighting approach}: The third approach to personalize a model is by increasing the weights in the loss function during the training process for speech samples in the adaptation set (i.e., same set used in the unique speaker approach). Unlike the previous two approaches, which adapt a pre-trained system, this approach trains the regression model from scratch. {\color{black}As described in Equation \ref{eq:loss}, we use $\mathcal{L}=(1 - CCC)$ as the loss function to train our models. For the weighting approach, this cost is assigned to a sample in the train set, but not on the adaptation set. For samples in the adaptation set, we multiply the cost $\mathcal{L}$ by a factor $\lambda > 1$. Therefore, an error of a sample from the adaptation set is $\lambda$ times more costly than an error made on other samples from the train set not included in the adaptation set.} We experiment with weighting ratios of 1:2, 1:3, 1:4 and 1:5, where higher weights are assigned to samples in the adaptation set. This approach uses all the train set, increasing the importance of correctly predicting samples in the adaptation set.

The proposed unsupervised adaptation schemes can be jointly applied to all the speakers in the test set, creating a single model. We refer to this approach as \emph{global adaptation} (GA) model. Alternatively, the approaches can be individually implemented for each speaker, creating as many models as speakers in the test set. This implementation only works for the unique speaker and weighting approaches. The oversampling approach does not apply in this case, since each speech segment in the adaptation set is drawn only once (i.e., we consider one test speaker at a time). We refer to this approach as the \emph{individual adaptation} (IA) model. We evaluate both implementations in Section \ref{sec:results} using adaptation sets of different sizes. 

\vspace{-0.3cm}
\section{Experimental Results}
\label{sec:results}

The prediction of valence is formulated as a regression problem implemented with DNNs with four dense layers and 512 nodes per layer. {\color{black} This setting achieved the best performance for valence on Table \ref{tab:results_within2}.} We use \emph{rectified linear unit} (ReLU) at the hidden layers and linear activations for the output layer. The DNNs are trained with batch normalization for the hidden layers. We use a dropout rate of $p=0.7$ at the hidden layers. The selection of this rate follows the findings in Section \ref{ssec:dropout_rate}, which demonstrates that a higher value of dropout is important for improving the detection of valence. We pre-train the models for 200 epochs with an early stopping criterion based on the performance on the development set. The best model is used to evaluate the results. The DNNs are trained with \emph{stochastic gradient descent} (SGD) with momentum of 0.9, and a learning rate of $r=0.001$. For the unique speaker and oversampling approaches, the learning rate is reduced to $r_{\mathit{adap}}=0.0001$ while adapting the regression model. We adapt the models with these approaches for 100 extra epochs with an early stopping criterion based on the performance on the development set. For the weighting approach, we train the models from scratch for 200 epochs with early stopping criterion, maximizing the performance on the development set. The loss function (Eq. \ref{eq:loss}) relies on the \emph{concordance correlation coefficient} (CCC), which is defined in Equation \ref{eq:ccc}.

\vspace{-0.3cm}
\begin{eqnarray}
\mathcal{L} &=&(1 - CCC) \label{eq:loss}\\
\mathit{CCC} &=& \frac{2\rho\sigma_x\sigma_y}{\sigma_x^2+\sigma_y^2+(\mu_x - \mu_y)^2} \label{eq:ccc}
\end{eqnarray}
\vspace{-0.3cm}

The parameters $\mu_x$ and $\mu_y$, and $\sigma_x$ and $\sigma_y$ are the means and standard deviations of the true labels ($x$) and the predicted labels ($y$), and $\rho$ is the Pearson's correlation coefficient between them. CCC takes into account not only the correlation between the true emotional labels and their estimates, but also the difference in their means. This metric takes care of the bias-variance tradeoff when comparing the true and predicted labels. CCC is also the evaluation metric in all our experimental evaluation. 

The input to the DNNs is the 6,373D  acoustic feature vector (Sec. \ref{ssec:features}). The features are normalized to have zero mean and unit standard deviation. This normalization is done using the mean and standard deviation values estimated over the training samples. {\color{black}After this normalization, we expect the features to be within a reasonable range. We remove outliers by clipping the features with values that deviate more than three standard deviations from their means (i.e., $\mu_{f_i} - 3\sigma_{f_i} \leq f_i \leq \mu_{f_i} + 3\sigma_{f_i}$).} The output of the DNNs is the prediction score for the emotional attribute.

We use the speaker-independent and speaker-dependent models described in Section \ref{ssec:sp_dep} as baselines, where the results are listed in Table \ref{tab:results_within2}. The speaker-independent model does not rely on any adaptation scheme to personalize the models to perform better on the test set. As described in Section \ref{ssec:sp_dep}, the speaker-dependent model is built by adding the \emph{test-A} set to the train set, using partial information from the speakers. While this setting is not representative of the performance expected for the regression model when evaluated on speech from unknown speakers, it provides an upper bound performance to contextualize the improvements observed with our proposed personalization methods. For analysis purposes, we report the performance of the speaker-dependent models obtained with the addition of 50s, 100s, 150s, and 200s per speaker in the test set. These extra samples are obtained from the \emph{test-A} set. We consistently evaluate all the models using the \emph{test-B} set. This experimental setting creates fair comparisons, since these samples are not used to train any of the models. 

\vspace{-0.3cm}
\subsection{Global Adaptation Model}
\label{ssec:one_model}
We evaluate the performance of the system with the global adaptation model, where a single regression model is built. The three adaptation schemes are implemented by considering the 50 speakers in the test set. The adaptation set is obtained by identifying the closest speakers in the train set to each speaker in the test set (Sec. \ref{ssec:estimation_close_speakers}). We implement this approach by identifying the five closest speakers in the train set ($N=5$). Section \ref{ssec:NumberClosest} shows the results with different number of speakers. We incrementally add more samples by randomly selecting 50s, 100s, 150s, 200s, and 300s from the selected speakers associated with a given speaker in the test set (Sec. \ref{ssec:Personalization}). This process is repeated for each of the speakers in the test set to observe the performance trend as a function of the size of the adaptation data. 

\begin{figure}[tb]
	\centering
	\includegraphics[width=8cm]{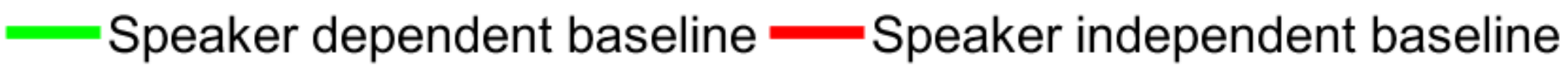}\\
	\includegraphics[width=7.5cm]{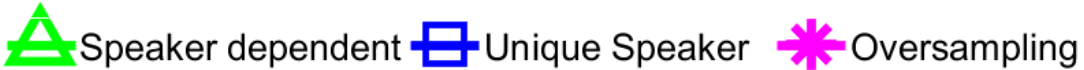}\\
	\includegraphics[width=0.9\columnwidth]{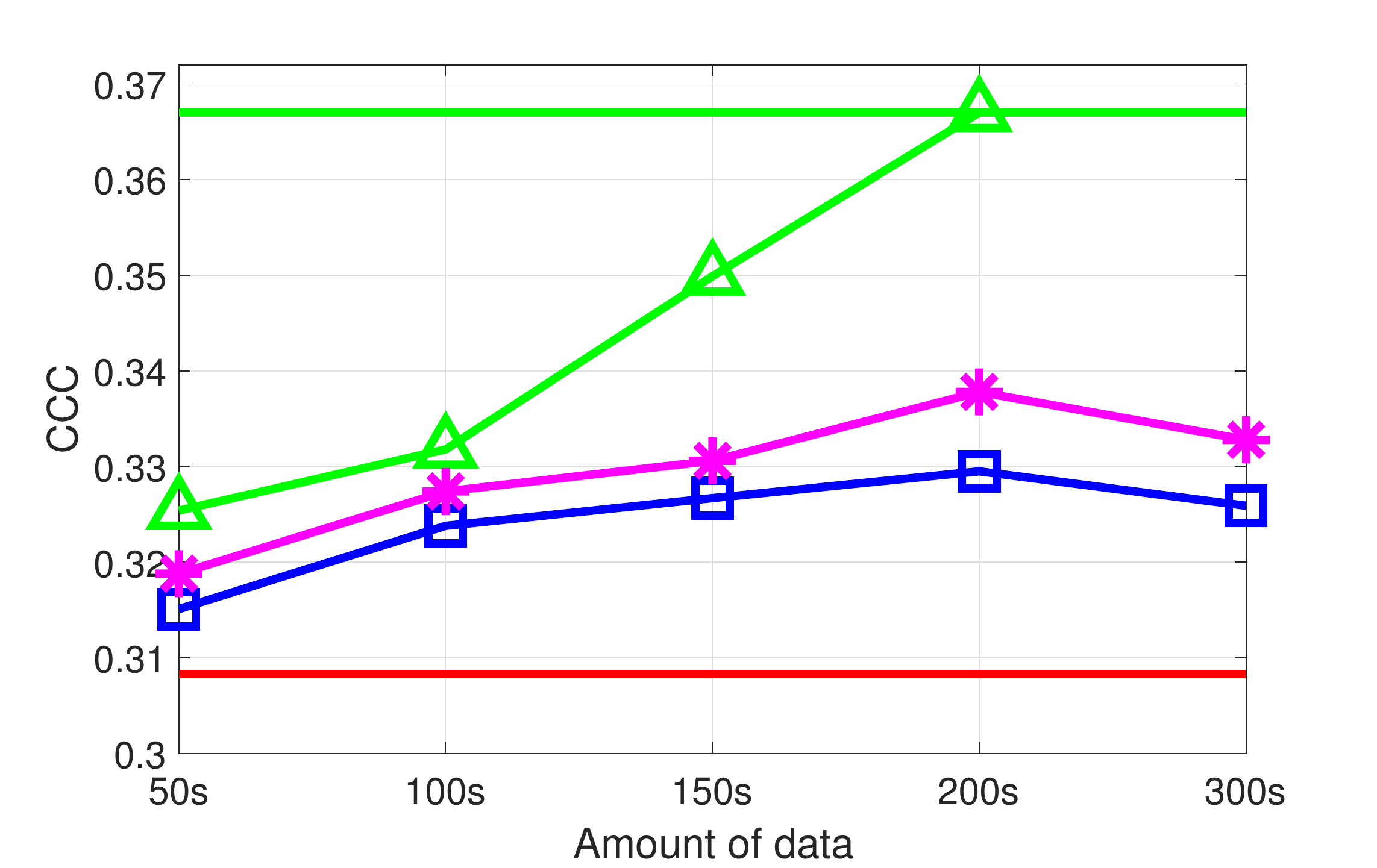}
	\caption{Results on the \emph{test-B} set for the global adaptation model for the unique speaker and oversampling methods. It shows the results for the speaker-dependent and speaker-independent baselines. It also shows the speaker-dependent model implemented with different size of data from the  \emph{test-A} set.}
	\label{fig:adapt1}
	\vspace{-0.3cm}
\end{figure}

First, we evaluate the unique speaker and oversampling approaches, which rely on model adaptation. Figure \ref{fig:adapt1} shows the results. The two solid horizontal lines are the speaker-dependent (green) and speaker-independent (red) baselines. The green line (triangle) corresponds to the speaker-dependent model as we increase the amount of data from the \emph{test-A} set. The performance for the unique speaker approach is shown in blue (square).  We clearly observe an improvement over the speaker-independent baseline, which demonstrate that the adaptation scheme is effective even with a very small adaptation set (e.g., 50s). The pink (asterisk) line in Figure \ref{fig:adapt1} shows the performance of the oversampling approach, which leads to better performance than the unique speaker approach. {\color{black}Both approaches use the same amount of adaptation data, but rely on different criteria to select the adaptation samples. Adding samples in multiple mini-batches according to the oversampling strategy is beneficial to improve the prediction of valence. For both models, we observe consistent improvement in CCC as more data is added into the adaptation set from 50s to 200s. After this point, the performance seems to saturate, observing fluctuations. Interestingly, adaptation with 200s of data using the oversampling approach leads to better performance than the speaker-dependent baseline implemented with 50s and 100s of data from the \emph{test-A} set.}

\begin{figure}[tb]
	\centering
	\includegraphics[width=8cm]{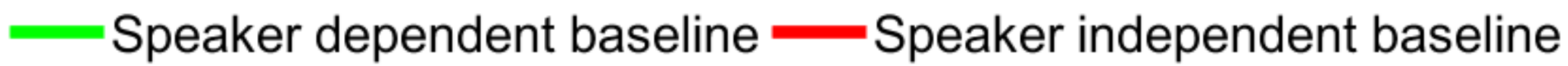}
	\includegraphics[width=8cm]{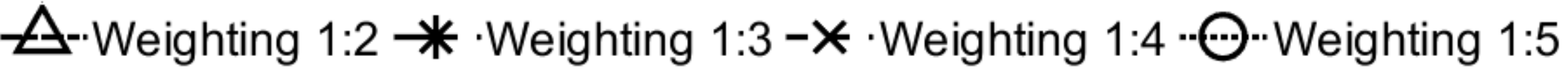}
	\includegraphics[width=0.9\columnwidth]{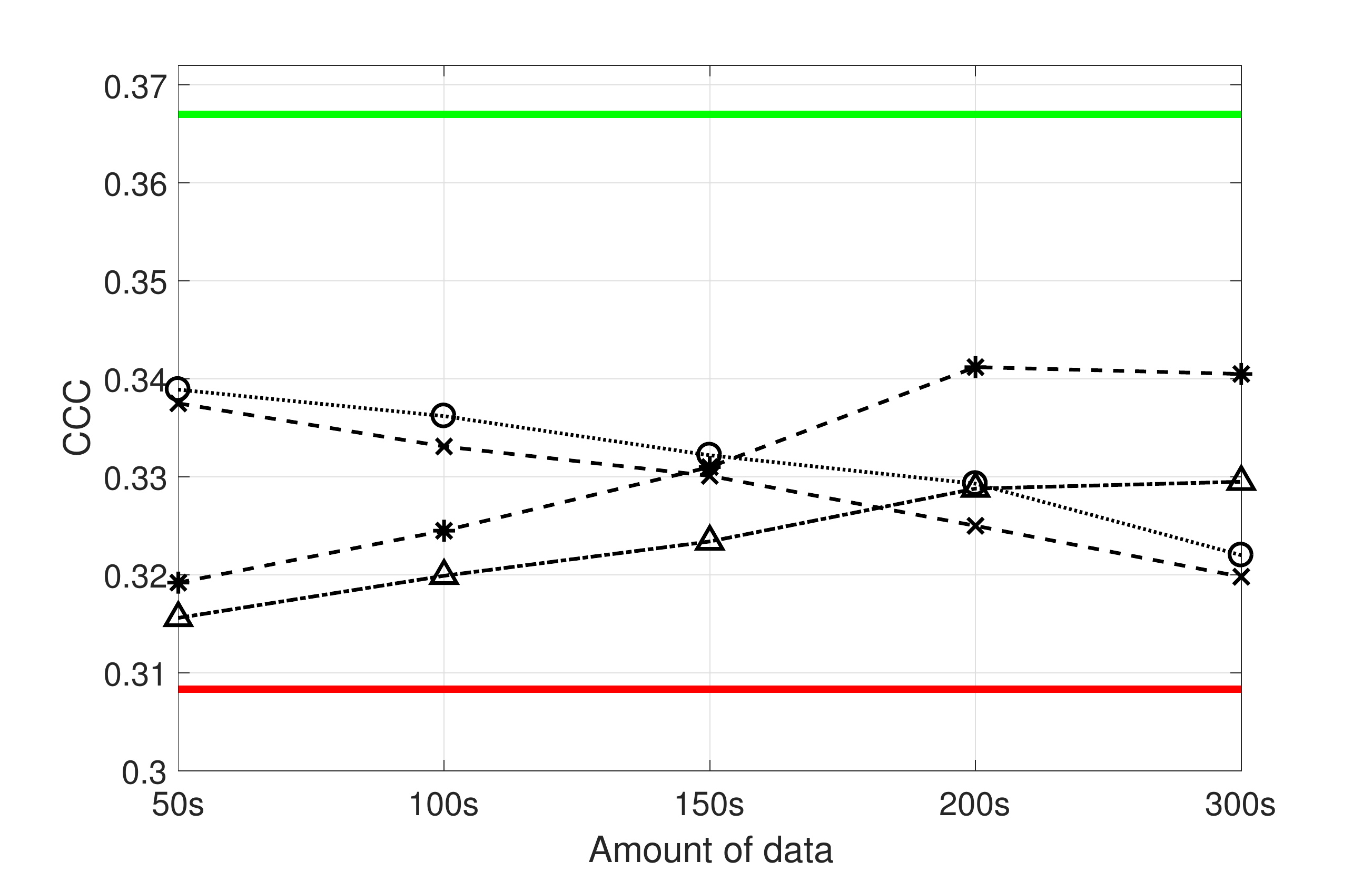}
    \caption{Results on the \emph{test-B} set for the global adaptation model for the weighting approach. The figure shows the results for the speaker-dependent and speaker-independent baselines.}
	\label{fig:adapt3}
	\vspace{-0.3cm}
\end{figure}

Second, we evaluate the performance of the weighting method, which trains the models from scratch, weighting more the samples in the adaptation set (Sec. \ref{ssec:Personalization}). We evaluate the amount of data included in this selected set, including 50s, 100s, 150s, 200s per speaker in the test set. We also consider using all the data from the selected speakers. Only samples in this set are weighed more, implementing this approach with different ratios (1:2, 1:3, 1:4 and 1:5). Figure \ref{fig:adapt3} shows the results. For weighting ratios 1:2 and 1:3, the performance gradually increases by adding more data in the selected set, peaking at 200s per speaker. However, the opposite trend is observed when the weighting ratios are either 1:4 or 1:5. Increasing the weights of speech samples in the adaptation set  diminishes the information provided in the rest of the train data, leading to worse performance. The best performance is obtained with a weighting ratio of 1:3, when the selected set includes 200s per speaker. Figures \ref{fig:adapt1} and \ref{fig:adapt3} show that this setting achieves similar performance than the best setting for the oversampling approach. 

\vspace{-0.3cm}
\subsection{Individual Adaptation Model}
\label{ssec:50_models}

This section presents the results of our approach implemented using the individual adaptation model. This approach builds one model for each of the speakers in the test set, creating the adaptation set with the samples from the speakers in the train set that are closer to this speaker (i.e., 50 separate models). {\color{black}For each model, we attempt to select equal duration of speech samples from each of the closest train speakers in the adaptation set to balance the amount of data used from each speaker.} After adapting the models, the results are reported by concatenating the predicted vectors for each speaker in the test set. We estimate the CCC values for the entire \emph{test-B} set. The approach is implemented with the five closest speakers to each speaker in the test set. The performance of the approaches are reported by increasing the adaptation set. As explained in Section \ref{ssec:Personalization}, we only evaluate the unique speaker and weighting approaches, since the oversampling approach cannot be implemented with a single speaker in the test set. 

\begin{figure}[tb]
	\centering
	\includegraphics[width=8cm]{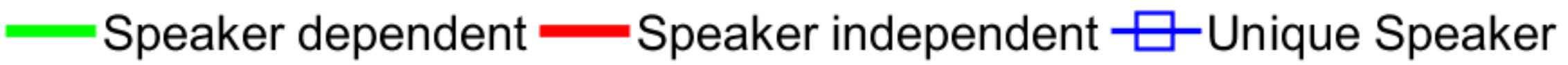}
	\includegraphics[width=8cm]{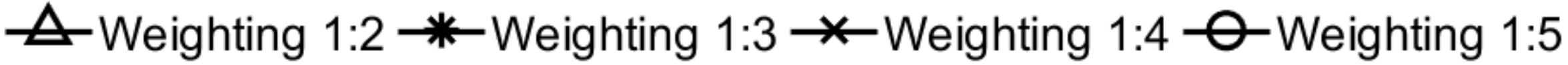}
	\includegraphics[width=1\columnwidth]{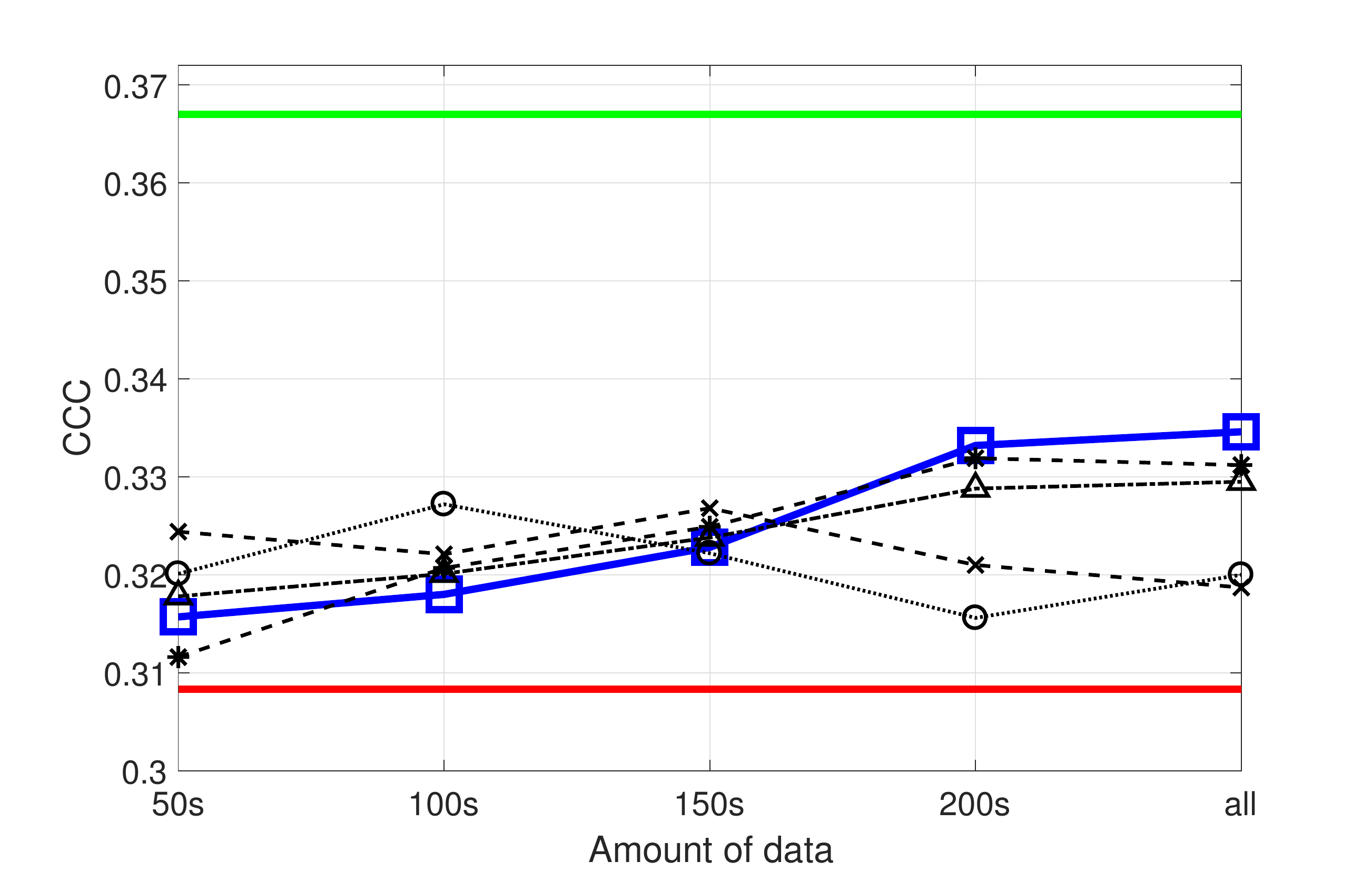}
	\caption{Results on the \emph{test-B} set for the individual adaptation model using the unique speaker and weighting approaches. The figure shows the results for the speaker-dependent and speaker-independent baselines.}
	\label{fig:adapt2}
	\vspace{-0.3cm}
\end{figure}

Figure \ref{fig:adapt2} shows the CCC scores obtained for different sizes of the adaptation set. The results show improvements over the speaker-independent baseline performance. The performance gains are consistently higher when all the data from the closest speakers are used. The weighting approach also leads to better performance than the unique speaker approach. However, the results are worse than the CCC values of approaches implemented with the global adaptation model (Figs. \ref{fig:adapt1} and \ref{fig:adapt3}). {\color{black}The decrease in performance in the IA model can be associated with the adaptation procedure. In the IA models, we adapt separate models for each target speaker. This procedure involves adapting 50 different models where their parameters and hyperplanes change  based on a small adaptation set. This approach may be too aggressive, resulting in lower performance than adapting a single model, considering all the target speakers. We have seen similar observations in the area of active learning in SER tasks, where a more conservative adaptation strategy led to better results  \cite{Abdelwahab_2017_2}. In the GA models, we are adapting a single model, where the shape and direction of the change of the hyperplane are smoother than in the case of the IA models achieving a better and stable performance}.

\vspace{-0.3cm}
\subsection{Number of Closest Speakers}
\label{ssec:NumberClosest}

\begin{figure}[tb]
	\centering
	\includegraphics[width=0.9\columnwidth]{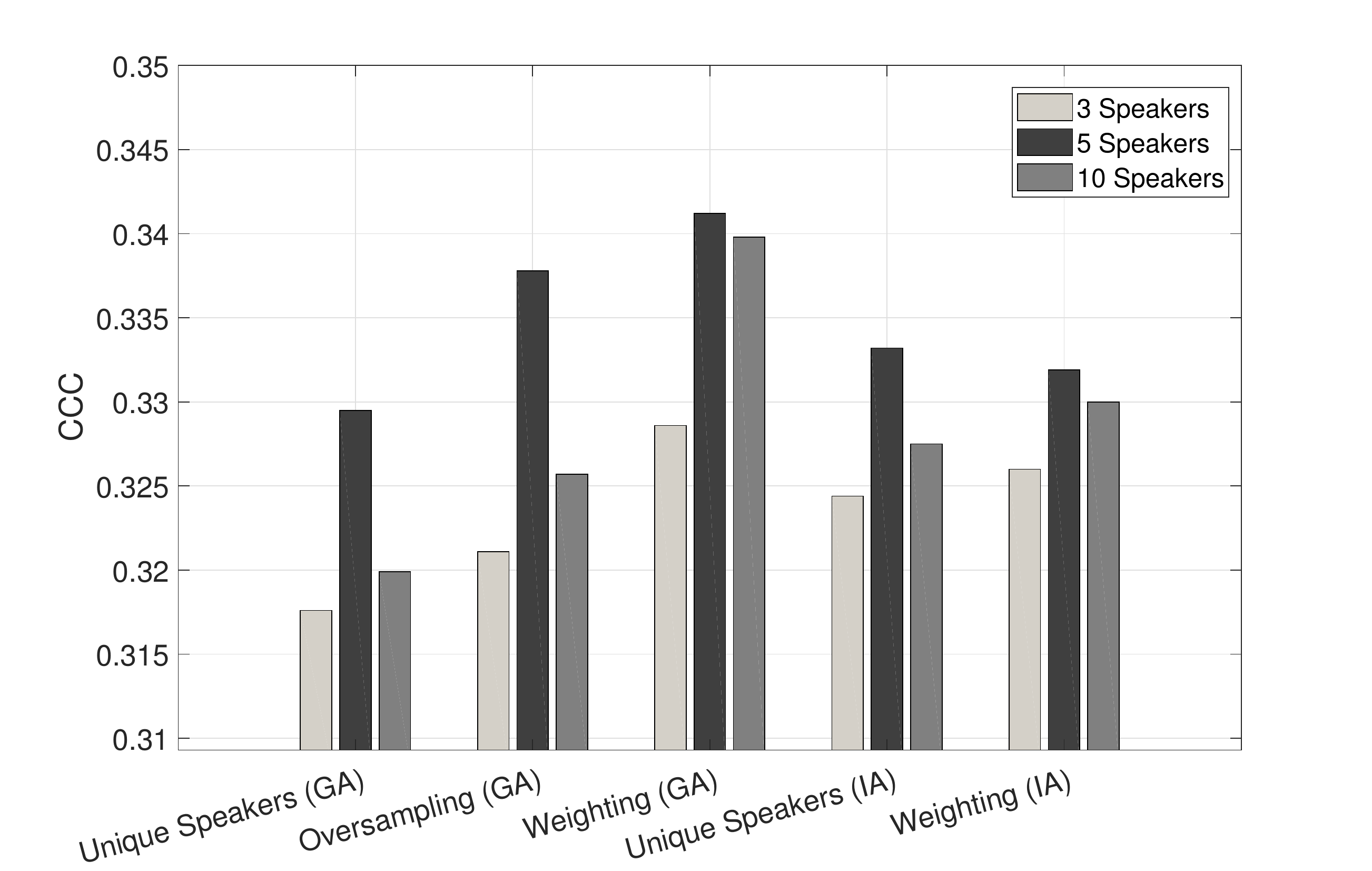}
	\caption{Evaluation of the optimal number of closer speakers ($N$) from the train set for each of the speakers in the test set. The training data from the selected speakers is included in the adaptation set. The results corresponds to the CCC values obtained with different methods on the \emph{test-B} set.}
	\label{fig:Closest}
\end{figure}

This section evaluates the number of closest speakers ($N$) from the train set selected for each speaker in the test set. If this number is too small, the adaptation set will not have enough variability. If this number is too high, we will select speakers that are not very close to the target speakers. We implement the weighting approach with the ratio 1:3. Figure \ref{fig:Closest} shows the results on the \emph{test-B} set for the models implemented with the global and individual adaptation models using the proposed adaptation schemes ($N\in \{3,5,10\}$). The results clearly show that $N=5$ is a good balance, obtaining higher performance across conditions. We set $N=5$ for the rest of the experimental evaluation. 


\vspace{-0.3cm}
\subsection{Minimizing Loss on Adaptation Set}
\label{ssec:training_loss}


\begin{figure}[t]
	\centering
	\includegraphics[width=5.5cm]{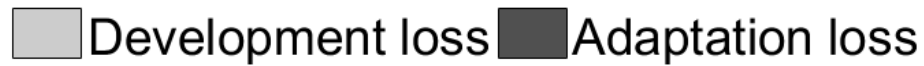}
	\subfigure[Global adaptation model]{
		\includegraphics[trim=0 0.5cm 0 0, clip,width=0.9\columnwidth]{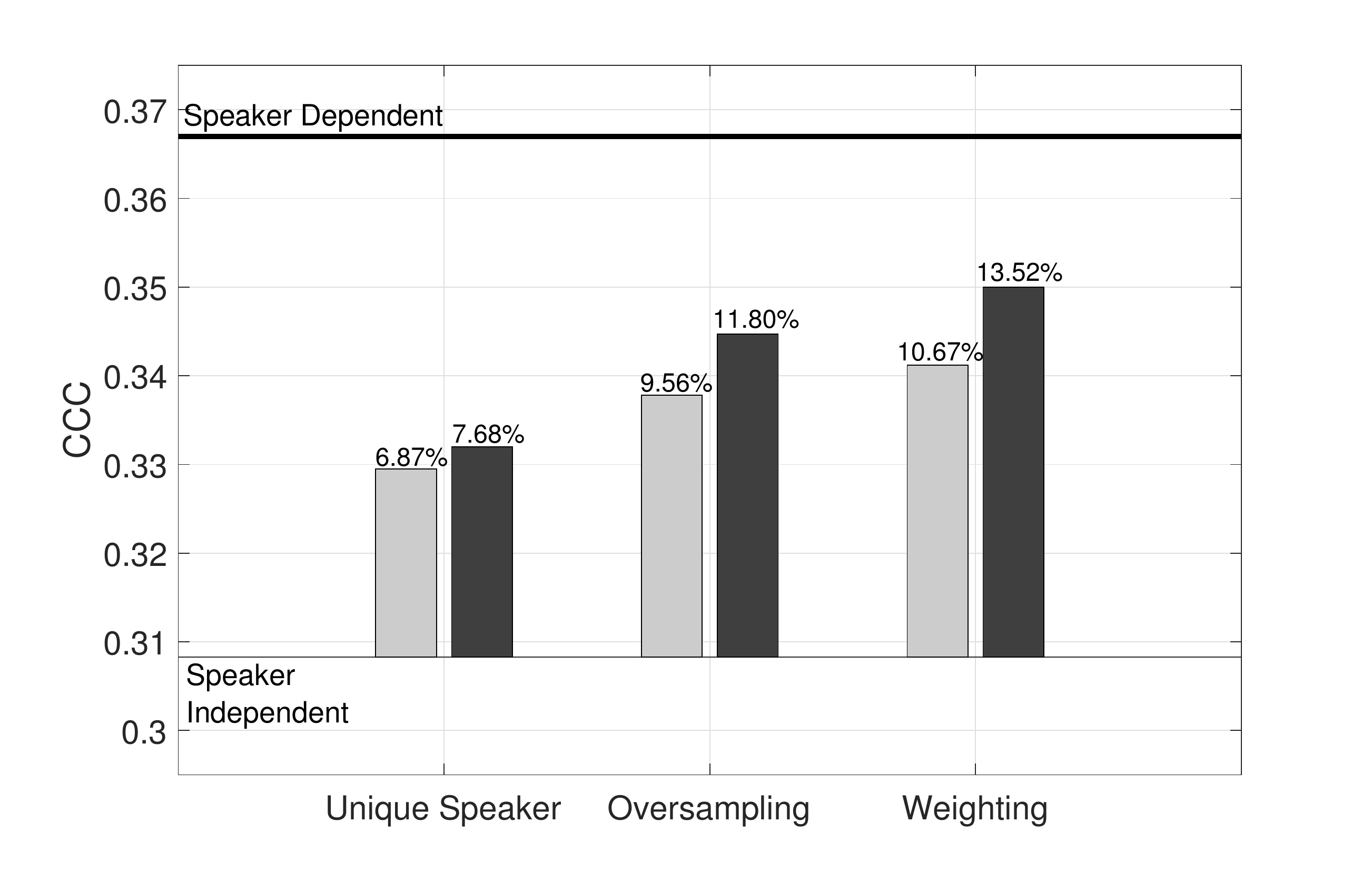}
		\label{fig:tr_loss_a}
	}
	\subfigure[Individual adaptation model]{
		\includegraphics[trim=0 0 0 0, clip,width=0.9\columnwidth]{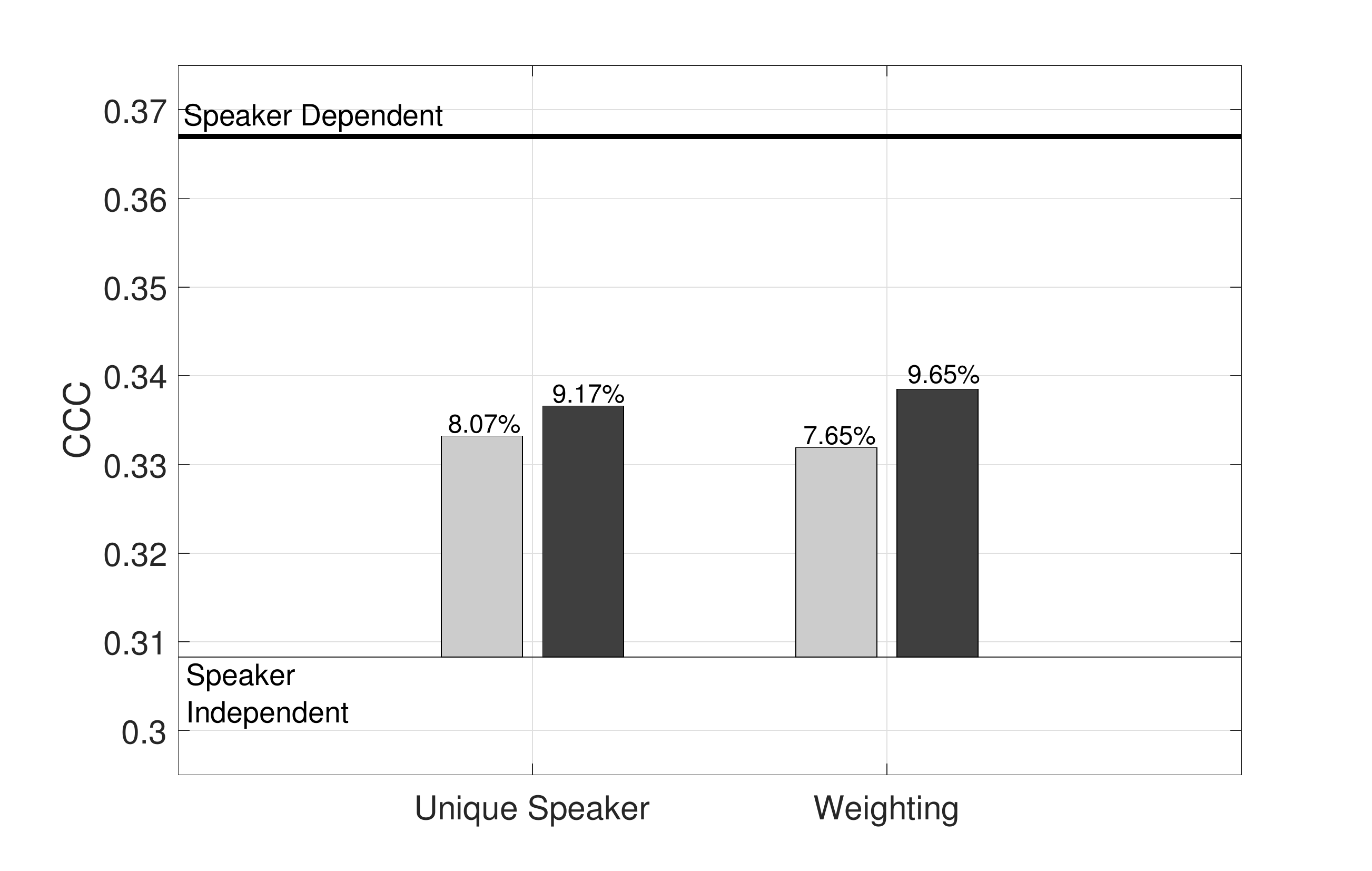}
		\label{fig:tr_loss_b}
	}
	\caption{Improvement in performance achieved by monitoring the loss function on the adaptation set while fine-tuning the models. The percentages over the bars indicate the relative improvements over the speaker-independent baseline. The figure shows the performance observed in the \emph{test-B} set.}
	\vspace{-0.4cm}
	\label{fig:tr_loss}
\end{figure}

The results of the model adaptation presented in Sections \ref{ssec:one_model}, \ref{ssec:50_models} and \ref{ssec:NumberClosest} are obtained by minimizing the loss function on the development set, a practice that aims to increase the generalization of the models. In our formulation, however, we aim to personalize a model towards a known set of speakers in the test set. Given the assumption that the selected speakers in the train set are similar to the test set, we can optimize the system by minimizing the loss function on the adaptation set when fine-tuning the model (i.e., samples from the selected speakers used for adaptation). Since we want to \emph{personalize} the system, it is theoretically correct to maximize the performance of the model on data that is found to be closer to the target speakers. This section evaluates this idea. We record the best performance of the model by using an early stopping criterion on the adaptation loss. The only special case is the weighting approach which trains the models from scratch. Since monitoring the loss exclusively on the adaptation set will ignore other samples in the train set, we decide to monitor the loss on the full train set. The differences in the weights increase the emphasis of samples from the adaptation set, achieving essentially the same goal. We use the adaptation set using the 200s condition. The weighting approach is implemented with the 1:3 ratio, which gave the best CCC in previous experiments (Figs. \ref{fig:adapt3} and \ref{fig:adapt2}).

Figure \ref{fig:tr_loss} shows the performance improvements achieved by minimizing the loss function on the adaptation set. The darker bars indicate the results obtained while monitoring the training loss,  and the lighter bars indicate the results obtained while monitoring the development loss. We include the relative improvements over the speaker-independent baseline with numbers on top of the bars. The relative improvements are constantly higher when maximizing the performance on the train set, tailoring even more the models to the test speakers. We conducted a one-tailed student t-test over ten trials to assert if the differences in performance between minimizing the loss function on the development and train sets are statistically significant. We assert significance when $p$-values$<$0.05. The statistical test indicates that the differences are statistically significant for  all the adaptation approaches implemented with either the global or individual adaptation models. This approach leads to performances that are closer to the speaker-dependent baseline.

\vspace{-0.3cm}
\subsection{Performance on Other Emotional Attributes}
\label{ssec:comparison}

The premise of this study is that valence is externalized in speech with more speaker-dependent traits than arousal and dominance. Therefore, personalization approaches to bring the models closer to the test speakers should have a higher impact on valence. This section implements the proposed adaptation schemes on arousal and dominance, comparing the relative improvements over the speaker-independent baseline with the results for valence.

\begin{table}[t]		
	\caption{Performance achieved using different adaptation approaches on the \emph{test-B} set. The table reports the performance gain over the speaker-independent baseline reported in Table \ref{tab:results_within2}}
	\centering
	\fontsize{8}{9}\selectfont
	\begin{tabular*}{1\columnwidth}{@{\extracolsep{\fill}}c|l|cc|cc}
		\hline
		 & \multicolumn{1}{c|}{Adaptation} &\multicolumn{2}{c|}{Minimizing loss} & \multicolumn{2}{c}{Minimizing loss}\\
		& \multicolumn{1}{c|}{Scheme} &\multicolumn{2}{c|}{in development set} & \multicolumn{2}{c}{in adaptation set}\\
		\cline{3-6}
		&&CCC & Gain (\%)& CCC & Gain (\%)\\
		\hline
		\hline
		\multirow{5}{*}{\rotatebox{90}{Valence}}
		&Unique speaker (GA)& 0.3295 & 6.87  & 0.3320 & 7.68\\
        &Oversampling (GA)& 0.3378 & 9.56 & 0.3447 & 11.80\\
        &Weighting (GA)& \textbf{0.3412} & \textbf{10.67} & \textbf{0.3500} & \textbf{13.52}\\
        &Unique speaker (IA)& 0.3332 & 8.07 & 0.3366 & 9.17\\
        &Weighting (IA)& 0.3319 & 7.65 & 0.3385 & 9.79\\
		\hline
		\multirow{5}{*}{\rotatebox{90}{Arousal}}
		&Unique speaker (GA)& 0.7196 & 0.44 & 0.7221 & 0.79\\
        &Oversampling (GA)& 0.7209 & 0.62 & 0.7258 & 1.31\\
        &Weighting (GA)& \textbf{0.7222} & \textbf{0.80} & \textbf{0.7296} & \textbf{1.84}\\
        &Unique speaker (IA)& 0.7185 & 0.29 & 0.7267 & 1.43\\
        &Weighting (IA)& 0.7202 & 0.53 & 0.7271& 1.49\\
		\hline
		\multirow{5}{*}{\rotatebox{90}{Dominance}}
		&Unique speaker (GA)& 0.6410 & 0.56 & 0.6415 & 0.64\\
        &Oversampling (GA)& 0.6428 & 0.84 & 0.6430 & 0.87\\
        &Weighting (GA)& \textbf{0.6433} & \textbf{0.92} & \textbf{0.6451} & \textbf{1.20}\\
        &Unique speaker (IA)& 0.6399 & 0.39 & 0.6419 & 0.70\\
        &Weighting (IA)& 0.6417 & 0.67 & 0.6422 & 0.75\\
		\hline
	\end{tabular*}
	\label{tab:results_all_version6}
	\vspace{-0.3cm}
\end{table}

Table \ref{tab:results_all_version6} reports the performance and relative improvements over the speaker-independent baseline for valence, arousal, and dominance when using the proposed methods. {\color{black}The relative improvements for arousal and dominance are less than 1.9\%, mirroring the results observed in Table \ref{tab:results_within2} that shows relative improvements less than 3\% for arousal and dominance when labeled data from the target speakers is available (i.e., speaker-dependent condition). Therefore, it is not surprising that the method does not lead to big improvements for arousal and dominance where there is little room for improvements. We argue that the approach is successful even for arousal or dominance, since the relative improvements for these emotional attributes are similar to the values reported in Table 1 under speaker dependent conditions.} In contrast, the relative improvements for valence are as high as 13.52\%. These results validate our hypothesis that exploiting speaker-dependent characteristics between train and test speakers helps to personalize a SER system in the prediction of valence.

\vspace{-0.3cm}
{\color{black}
\subsection{Comparison with Other Baselines}
\label{ssec:baselines}

\begin{table}[t]		
	\caption{Comparison of results in terms CCC obtained with the proposed personalization approach and other methods. All the experiments are done with MSP-Podcast corpus and evaluated on Test-B set. STL: Single Task Learning (Speaker-Independent baseline), MTL: Multi-Task Learning}
	\centering
	\fontsize{8}{9}\selectfont
	\begin{tabular*}{0.98\columnwidth}{@{\extracolsep{\fill}}c|c|c|c|c}
		\hline
		 Attributes & Proposed & STL & MTL & Ladder\\
		 &Approach&&&Networks\\
		\hline
		\multirow{1}{*}{Valence}
        & \textbf{0.3500} & 0.3083 & 0.3302 & 0.3158\\
		\hline
        \multirow{1}{*}{Arousal}
        & 0.7296 & 0.7164 & 0.7214 & \textbf{0.7421}\\
		\hline
		\multirow{1}{*}{Dominance}
        & 0.6451 & 0.6374 & 0.6287 & \textbf{0.6498}\\
		\hline
	\end{tabular*}
	\label{tab:comparison_results}
	\vspace{-0.3cm}
\end{table}

We compare the results from our proposed personalization approach with other state-of-the art approaches. We consider \emph{multi-task learning} (MTL) \cite{Parthasarathy_2017_3}, and ladder network \cite{Parthasarathy_2018_3} for SER. These are some of the most successful approaches used in SER. The MTL approach jointly predicts arousal, valence and dominance, where the loss function is a weighted sum of the individual attribute losses. We used $\mathcal{L} = (1 - CCC)$ as the loss for each attribute ($\mathcal{L}_{aro}$, $\mathcal{L}_{val}$, $\mathcal{L}_{dom}$). Equation \ref{eq:mtl}  shows the overall loss function, where $(\alpha,\beta) \in [0, 1]$ and $\alpha + \beta \leq 1$. The hyperparameters $\alpha$ and $\beta$ are tuned on the development set.

\vspace{-0.3cm}
\begin{eqnarray}
\mathcal{L}_{MTL} = \alpha \mathcal{L}_{aro} + \beta \mathcal{L}_{val} + (1 - \alpha - \beta) \mathcal{L}_{dom} 
\label{eq:mtl}
\end{eqnarray}

The ladder network approach follows the  implementation presented by Parthasarathy and Busso \cite{Parthasarathy_2018_3}. This method uses the reconstruction of feature representations at various layers in a DNN as auxiliary tasks. In addition, we consider the speaker-independent baseline discussed in previous sections, referred here to as \emph{single task learning} (STL). 

Table \ref{tab:comparison_results} describes the results, which clearly show significant improvements over alternative methods in CCC for valence using our proposed approach, reinforcing our claim about the benefits of personalization in the estimation of valence. Other approaches are effective for arousal and dominance.

\vspace{-0.3cm}
\subsection{Performance on Other Corpora}
\label{ssec:iemocap_improv}

\begin{table}[t]		
	\caption{Comparison of CCC values between speaker-independent and speaker-dependent conditions for IEMOCAP (IEM) and MSP-IMPROV (IMP) databases. The DNN is trained with three layers and $256$ nodes per layer. The column `Gain' shows the relative improvement by training with partial data from the target speakers (\emph{test-A} set).}
	\centering
	\fontsize{8}{9}\selectfont
	\begin{tabular*}{0.98\columnwidth}{@{\extracolsep{\fill}}c|c|c|c|c}
		\hline
		 Attributes & Database & Speaker & Speaker & Gain\\
		 &&Independent&Dependent\\
		 \cline{3-5}
		 && \emph{Test-B} set& \emph{Test-B} set& (\%)\\
		\hline
		\hline
		\multirow{2}{*}{Valence}
        & IEM & 0.4428 & 0.5072 & 14.54\\
		& IMP & 0.3420 & 0.4164 & 21.75\\        
		\hline
        \multirow{2}{*}{Arousal}
        & IEM & 0.6953 & 0.7255 & 4.34\\
        & IMP & 0.5958 & 0.6218 & 4.36\\
		\hline
		\multirow{2}{*}{Dominance}
        & IEM & 0.5444 & 0.5678 & 4.29\\
        & IMP & 0.5625 & 0.4911 & 6.18\\
		\hline
	\end{tabular*}
	\label{tab:results_within_iemcap_improv}
\end{table}

\begin{table}[t]		
	\caption{IEMOCAP and MSP-IMPROV: Performance achieved using different adaptation approaches on the \emph{test-B} set. The table reports the performance gain over the speaker-independent baselines reported in Table \ref{tab:results_within_iemcap_improv}. All the experimental evaluations are done by minimizing the loss in the development set.}
	\centering
	\fontsize{8}{9}\selectfont
	\begin{tabular*}{1\columnwidth}{@{\extracolsep{\fill}}c|l|cc|cc}
		\hline
		 & \multicolumn{1}{c|}{Adaptation} &\multicolumn{2}{c|}{USC-IEMOCAP} & \multicolumn{2}{c}{MSP-IMPROV}\\
		& \multicolumn{1}{c|}{Scheme} & & \\
		\cline{3-6}
		&&CCC & Gain (\%)& CCC & Gain (\%)\\
		\hline
		\hline
		\multirow{5}{*}{\rotatebox{90}{Valence}}
		&Unique speaker (GA)& 0.4761 & 7.52  & 0.3866 & 13.04\\
        &Oversampling (GA)& 0.4790 & 8.17 & 0.4014 & 17.36\\
        &Weighting (GA)& 0.4889 & 10.41 & 0.3915 & 15.52\\
        &Unique speaker (IA)& 0.4725 & 6.70 & 0.3855 & 12.71\\
        &Weighting (IA)& 0.4759 & 7.47 & 0.3873 & 13.24\\
		\hline
		\multirow{5}{*}{\rotatebox{90}{Arousal}}
		&Unique speaker (GA)& 0.6988 & 0.50 & 0.6057 & 1.66\\
        &Oversampling (GA)& 0.7001 & 0.69 & 0.6086 & 2.14\\
        &Weighting (GA)& 0.7002 & 0.70 & 0.6191 & 3.91\\
        &Unique speaker (IA)& 0.6966 & 0.18 & 0.6087 & 2.16\\
        &Weighting (IA)& 0.7000 & 0.68 & 0.6095 & 2.29\\
		\hline
		\multirow{5}{*}{\rotatebox{90}{Dominance}}
		&Unique speaker (GA)& 0.5491 & 0.86 & 0.4718 & 2.01\\
        &Oversampling (GA)& 0.5500 & 1.02 & 0.4731 & 2.29\\
        &Weighting (GA)& 0.5495 & 0.93 & 0.4820 & 4.21\\
        &Unique speaker (IA)& 0.5451 & 0.12 & 0.4710 & 1.83\\
        &Weighting (IA)& 0.5496 & 0.95 & 0.4713 & 1.90\\
		\hline
	\end{tabular*}
	\label{tab:results_all_iemocap_improv}
	\vspace{-0.3cm}
\end{table}

We validate the effectiveness of the proposed personalization approaches with other emotional databases. We use a $K$-fold cross-validation strategy to train DNN models using the USC-IEMOCAP and MSP-IMPROV databases. In each fold, we consider $2$ speakers as the test speakers and the rest of the speakers as the train speakers. With this cross-validation approach, all the speakers are at some point considered as the test speakers. The final results are averaged across the $K$ folds. Similar to Figure \ref{fig:corpus}, we split the test set of both IEMOCAP and MSP-IMPROV databases into two partitions for this study: \emph{test-A} and \emph{test-B} sets. The \emph{test-A} set includes 200s of recording for each of the speakers in the test set. This set is reserved for finding the closest training speakers to the target speakers. The \emph{test-B} set includes the rest of the recordings in the test set. We implement the global and individual adaptation models with all the different adaptation approaches by minimizing the loss in the development set.

Table \ref{tab:results_within_iemcap_improv} shows the CCC results for speaker-independent and speaker-dependent conditions using the USC-IEMOCAP and the MSP-IMPROV corpora. The results validate the findings in Table \ref{tab:results_within2}, showing that the speaker-dependent condition leads to higher relative gains for valence than for arousal or dominance. 

Table \ref{tab:results_all_iemocap_improv} shows the results obtained with different adaptation schemes on the USC-IEMOCAP and MSP-IMPROV databases. The results are consistent with the findings observed with the MSP-Podcast corpus. We observe that the relative improvements over the speaker-independent baseline are much higher for valence than the relative improvements for arousal and dominance. With the USC-IEMOCAP corpus, we achieve relative gains in performance up to 10.41\% for valence whereas less than 1.03\% for arousal and dominance. Similarly, with the MSP-IMPROV corpus, we achieve relative gains in performance up to 17.36\% for valence whereas less than 4.22\% for arousal and dominance. These results show the effectiveness of our proposed approach applied to  other emotional corpora, reinforcing our finding about the speaker-dependent nature of valence emotional cues. 
}

\vspace{-0.3cm}
\section{Conclusions}
\label{sec:conclusion}

This paper demonstrated that a valence prediction system can be personalized to target speakers by exploiting speaker-dependent traits. The study proposed to create an adaptation set by identifying speakers in the train set that are closer to the speakers in the test set. Since we evaluate the similarity between speakers by comparing the acoustic feature spaces associated with each speaker without using emotional labels, the adaptation approaches are fully unsupervised. We proposed three methods to create this adaptation set: unique speaker, oversampling and weighting approaches. The adaptation sets from the selected \emph{closer} speakers are used to personalize the DNNs to the speakers in the test set. The experimental results showed that the global adaptation models achieved better performance than the individual adaptation models. Further improvements are observed when the loss function is minimized by monitoring the loss in the adaptation set. The proposed adaptation schemes lead to relative improvements up to 13.52\% over a speaker-independent baseline. We observed significant improvements in performance, even when only a few seconds of adaptation data (belonging to the train set) for each of the speakers in the test set was used for adaptation. However, increasing the amount of adaptation data did not contribute to further improvements of the model. The maximum performance gains were observed for 200s of adaptation data for each of the speakers in the test set. {\color{black}We also demonstrated the effectiveness of the proposed personalization approaches with the USC-IEMOCAP and MSP-IMPROV databases, showing consistent findings.}

{\color{black}
There are many interesting and important applications where our personalization models can be very helpful. In healthcare applications where medical data cannot be frequently obtained, a personalized system can keep track of a patient's expressive behaviors and his/her medical record for better and efficient treatment. Another example is on personal assistant systems on mobile devices, which are often used by a limited number of users. Over time, the system can collect enough data from the target users to improve their emotion recognition systems. For cloud-based applications, the training data needs to be stored in the cloud, instead of the edge device. With a simple modification on the approach to obtain the PCA projections (i.e., finding a common PCA space across testing speakers), the training data can be pre-estimated and stored. Therefore, this approach does not require storage or computational resources on the edge devices and can be efficiently implemented during inferences.}

This study demonstrated the importance of exploiting the speaker-dependent traits observed in the externalization of valence from speech, which led to clear improvements. It also showed that  we can personalize SER models by just finding speakers in the train set that are similar to the target speakers. The proposed formulation is flexible, requiring only to know the block of data associated with each speaker in the test set. This assumption is reasonable since it is straightforward to group data per speaker in many practical applications (e.g., assigning all the speech collected during a call center session to a single user). As a future work, we will evaluate more sophisticated methods to assess the similarity of speakers by considering more than acoustic similarities. Also, we will explore the use of the proposed adaptation schemes in other deep learning frameworks such as autoencoders, \emph{generative adversarial networks} (GANs) or \emph{long short-term memory} (LSTM). Another open question is to investigate adaptation schemes that are effective for arousal and dominance.

\appendices

\section*{Acknowledgment}

This study was funded by National Science Foundation (NSF) career award IIS-1453781

\ifCLASSOPTIONcaptionsoff
  \newpage
\fi


\bibliographystyle{IEEEtran}
\bibliography{reference}




\end{document}